\documentclass[aps,prd,groupedaddress,]{revtex4}
\usepackage{latexsym}
\usepackage[dvips]{graphicx}

\newcommand{\eq}{\begin{equation}}
\newcommand{\feq}{\end{equation}}
\newcommand{\eqn}{\begin{eqnarray}}
\newcommand{\feqn}{\end{eqnarray}}
\newcommand{\arr}{\begin{eqnarray*}}
\newcommand{\farr}{\end{eqnarray*}}
\newcommand{\beq}{\begin{equation}}
\newcommand{\eeq}{\end{equation}}
\newcommand{\bea}{\begin{eqnarray}}
\newcommand{\eea}{\end{eqnarray}}

\def\beq{\begin{equation}}
\def\eeq{\end{equation}}
\def\feq{\end{equation}}
\def\bea{\begin{eqnarray}}
\def\eea{\end{eqnarray}}
\def\bc{\begin{displaymath}}
\def\ec{\end{displaymath}}
\def\lb{\label}

\def\de{\delta}

\def\la{\lambda}

\def\lb{\label}

\begin{document}
\title{Acoustic analogues of black hole singularities}

\author{Mariano Cadoni}
\email{mariano.cadoni@ca.infn.it}

\affiliation{Dipartimento di Fisica,
Universit\`a di Cagliari, and INFN, Sezione di Cagliari, Cittadella
Universitaria 09042 Monserrato, Italy}
\author{Salvatore Mignemi}
\email{ smignemi@vaxca1.unica.it}
\affiliation{Dipartimento di  Matematica, Universit\`a di Cagliari,
viale Merello 92, 09123 Cagliari, Italy, and INFN, Sezione di Cagliari}


\begin{abstract}

We  search for  acoustic analogues of a spherical symmetric black
hole with a pointlike source. We show that the gravitational
system has a dynamical counterpart in the constrained, steady
motion of a fluid with a planar source. The equations governing
the dynamics of the gravitational system can be exactly mapped in
those governing the motion of the fluid. The different meaning
that singularities  and sources have in fluid dynamics and in
general relativity is also discussed. Whereas in the latter a
pointlike source is always associated with a (curvature)
singularity in the former the presence of  sources does not
necessarily imply divergences of the fields.

\end{abstract}


\maketitle


\maketitle

\section{Introduction}
One of the most intriguing features of classical general relativity is
the presence of curvature singularities. The classical
gravitational dynamics described by general relativity generates
singularities  both in the
collapse  of massive bodies and in the cosmological evolution of the
universe. The physical meaning of these  spacetime singularities has been widely
discussed since their discovery. By now it has become
conventional wisdom the point of view that the resolution of the
singularities of general relativity has to be found  in  a quantum theory of
gravity governing its short distance  behavior.
For instance,  string theory  allows for the quantum resolution
of  some singularities,  but
we are far away from having a detailed description of how the
Schwarzschild black hole (or the cosmological) singularity is
resolved.

Building on a proposal of Unruh \cite{Unruh:1980cg},
in recent years  there has been a flurry of activity about the
use of condensed matter systems  to mimic various
kinematical aspects of general relativity
\cite{Visser:1993ub,Visser:1997ux,Novello:2002qg,Visser:2001fe,Cardoso:2005ij}.
Condensed matter systems provide us with
analogue models of gravity that not only reproduce  typical
gravitational structures
(event and cosmological horizons) but can be also used to
indicate
an answer to fundamental questions of gravitational physics (field
theory in curved spacetime and Hawking radiation)
\cite{Volovik:2000ua, Barcelo:2001ca,fedichev0303,fedichev0304,
Barcelo:2003et,Volovik:2003ga, Barcelo:2004wz,Balbinot:2004da,Balbinot:2004dc,Berti:2004ju,
Cardoso:2004fi,Nakano:2004ha}.
The hope behind these efforts is that in a near future  we
could experimentally test the properties of the gravitational
interaction using  condensed matter systems analogues.

One is therefore tempted to  use analogue systems of gravity, in
particular fluids, to
try to mimic and understand spacetime singularities.
Till a few months ago this idea would have been
immediately rejected because the analogy gravity/fluids was
formulated only at a  kinematical level.
Spacetime singularities in general relativity have a dynamical origin.
In order to describe them using analogue models one needs to formulate
the correspondence at a dynamical level.

For spherically symmetric black holes such a dynamical correspondence
has been  recently found \cite{Cadoni:2004my}.
After gauge fixing, the four-dimensional dynamics governing
a spherically symmetric black hole was shown to be
equivalent to that governing the steady, constrained, cylindrically
symmetric  motion of a  fluid.
The results of Ref.\ \cite{Cadoni:2004my} provide us with a
framework to describe black hole singularities using acoustic analogues.

In this paper, extending the results of Ref.\ \cite{Cadoni:2004my}
we will find the dynamical acoustic analogue of a spherical symmetric
black hole with a pointlike source.
We will show that the gravitational system has a dynamical counterpart
in the constrained, steady motion of a fluid with a source (Sect.
II).
Although the discussion will be
purely classical it will shed light on the different
meaning that singularities  and sources have
in fluid dynamics and in general relativity. Whereas in the latter a
pointlike source is always associated with a (curvature)
singularity,
in the former the presence of sources does not necessarily imply
divergences of the fields.

We will also discuss the implications of our results for
black hole thermodynamics (Sect. III) and  solve the constraint for
the fluid dynamics in the case of a fluid with a
generic power-law equation of state (Sect. IV).

\section{Gravitational dynamics, pointlike sources and fluid dynamics}
We start from generic four-dimensional Einstein gravity coupled  to
matter fields and a  pointlike source of mass $m$.  The action
is (we adopt the notations of Ref.\ \cite{Wei} and   use natural
units $c=\hbar=k_{B}=1$)
\beq\lb{ea}
A=\frac{1}{16\pi G}\int d^4x\sqrt{-g}\ R- \int
d^4x\sqrt{-g}\ {\cal L}_{MF}- m\int
dt\sqrt{-g_{ij}\frac{dx^{i}}{dt}\frac{dx^{j}}{dt}},
\eeq
where  $G$ is the Newton constant and ${\cal L}_{MF}$ is
the Lagrangian for the matter fields.  ${\cal L}_{MF}$ not only  describes
matter fields such as  gauge and scalar fields,  but may also contain a
cosmological constant $\Lambda$. We shall consider  for
simplicity only the case
in which the classical solutions for the matter fields are  (or can be treated as)
constant background configurations.
In general the field equations stemming from the action (\ref{ea}) will
admit spherically symmetric black hole solutions, whose   mass $M$   will
be equal to the mass $m$ of the pointlike source. Moreover,
these solutions will exhibit curvature singularities at the location of the
source.  The simplest case is
represented by the Schwarzschild solution, obtained
setting ${\cal L}_{MF}=0$ in Eq.\ (\ref {ea}). Introducing a cosmological
constant term or an electromagnetic field in (\ref {ea}), we have
the Schwarzschild-anti de Sitter or the Reissner-Nordstrom solution, respectively.

We want to find a classical, non-relativistic fluid whose dynamics is
equivalent to that governing the static, spherically symmetric
solutions of the theory (\ref{ea}).
When the pointlike source is not present, this problem can be drastically
simplified by noticing that once
spherical symmetry is imposed the gravitational theory (\ref{ea}) becomes
essentially  two-dimensional (2D). Recently this result has been extended
also to the case where sources are present \cite{Cadoni:2005fw}.
The 2D model is obtained by retaining only the radial modes in
the  action (\ref{ea}) and it has the form of a 2D dilaton gravity
model (see for instance Ref.\ \cite{Grumiller:2002nm}).
A scalar field $\phi$ (the dilaton) parametrizes the volume of the transverse
$2$-dimensional sphere,
\beq\lb{an}
ds^{2}_{(4)}=ds^{2}_{(2)}+ \frac{2}{\lambda^{2}}\phi\,
d\Omega^{2}_{2},
\feq
where $\la$ is a parameter with dimensions of
length$^{-1}$,  related  to the 4D Newton constant, $G= \lambda^{-2}$.
Inserting the ansatz (\ref{an}) into the action (\ref{ea}) and
performing the Weyl rescaling of the 2D metric
\beq\lb{wr}
g\to \frac{1}{\sqrt{2\phi}}\ g,
\eeq
needed to get rid of the kinetic terms
for the dilaton, one obtains a 2D model characterized by a
dilaton potential $V(\phi)$  and a coupling function $W(\phi)$
\cite{Cadoni:2005fw},
\beq\lb{dg}
A=A_{G}+A_{M}= \frac{1}{2} \int d^{2}x\sqrt{-g}\left(\phi R
+\lambda^{2}V(\phi)\right)- m\int dt\, W(\phi)\sqrt{-g_{\mu\nu}\frac{dx^{\mu}}{dt}
\frac{dx^{\nu}}{dt}}.
\feq
The dilaton  potential  $V(\phi)$ depends on the form of the
matter Lagrangian ${\cal L}_{MF}$, whereas the coupling function $W(\phi)$
reads
\beq\lb{cf}
\quad W(\phi)=\left(
2\phi\right)^{-1/4}.
\feq

For static solutions in  the Schwarzschild gauge,
\beq\lb{gf}
ds^{2}= -
    U(r) d\tau^{2}
    + U^{-1}(r) dr^{2},
\feq
with a source particle at rest in the origin, the field equations
coming from the 2D action (\ref{dg}) become \cite{Cadoni:2005fw},
\bea\lb{fem}
&&\frac{d^{2}U}{dr^{2}}=\lambda^{2}\frac{dV}{d\phi}
-2m{dW\over d\phi}\,\de(r),\nonumber\\
&&\frac{dU}{dr}\frac{d\phi}{dr}=\lambda^{2}V,\\
&&2U\frac{d^{2}\phi}{dr^{2}}+\frac{dU}{dr}\frac{d\phi}{dr}=\lambda^{2}V
-2mW\,\de(r).\nonumber \eea
In order to find the acoustic analog of the  gravitational dynamics
(\ref{fem}) we will consider separately the gravitational dynamics in
vacuo, $m=0$,  and in the presence of the source
$m\neq0$.

\subsection{Gravitational dynamics in vacuo and fluid dynamics}

In this case the problem of finding a fluid dynamics  equivalent to
the gravitational dynamics (\ref{fem}) has been already solved in Ref.
\cite{Cadoni:2004my}. However,  in that paper the dilaton has not been
considered as dynamical, but rather as constrained by the equations of motion
to be proportional to the spacelike coordinate $r$ of the 2D spacetime.
Here we will  improve the derivation of Ref.\ \cite{Cadoni:2004my} by
considering the full gravitational dynamics with both degrees  of
freedom $U$ and $\phi$.

When  considering  the field equations (\ref{fem}) in the vacuum
or, equivalently, the spacetime region $r\neq0$  away from the source,
we can define a scalar mass function $M(r)$
\cite{Mann:1992yv}
\beq\lb{mass}
M(r)= \frac{F_{0}}{2}\left(\la^{2}\int V(\phi) d\phi -
(\nabla\phi)^{2}\right),
\feq
which is constant  by virtue of the field equations (\ref{fem}),
with $F_{0}$  arbitrary normalization constant.
On shell we therefore have $M(r)=M=const$, where  $M$ is the mass
associated with the  classical solution of the field equations under
consideration.

Using the mass function $M(r)$ we can find an equivalent form for the system
(\ref{fem}). Only two of the three equations in (\ref{fem})  are independent
and the the system  can be written, for $\phi\neq const$, as
\beq\lb{fe3}
\frac{dU}{dr} \frac{d\phi}{dr}=\lambda^{2}V,\quad\quad
\frac{dM(r)}{dr}=0.
\feq
By subtracting  the third from the second equation we can also rewrite
Eqs.\ ({\ref{fem}) in  a second  equivalent form,

\beq\lb{fe4}
\frac{dU}{dr} \frac{d\phi}{dr}=\lambda^{2}V,\quad\quad
\frac{d\phi^{2}}{dr^{2}}=0.
\feq
The existence of two equivalent forms  for the  equations  of the
gravitational  dynamics  can be
traced back to the existence of two constant of motion: $M$ and
$\phi_{0}=(1/\lambda)d\phi/dr$.
In the solution of Eqs.\ (\ref{fem}) these constants of motion appear as
integration constants,
\beq\lb{bh}
ds^{2}= -
 \phi_{0}^{-2}\left( J(\phi)- \frac{2M\phi_{0}}{\lambda}\right)d\tau^{2}
 + \phi_{0}^{2} \left( J(\phi)-
 \frac{2M\phi_{0}}{\lambda}\right)^{-1}dr^{2},\quad \phi=\phi_{0}\la r,
\feq
where $J=\int V d\phi $. Moreover,  they are related one to the other in a
geometrical
way:  $d\phi/dr$ determines the Killing vector of the metric
(\ref{bh}) $\chi^{\mu}= \epsilon ^{\mu\nu}\partial_{\nu}\phi$, whereas
$M$ is its associated conserved charge. In general the solution (\ref{bh})
describes a black hole with an event horizon at $r=r_{h}$, with
$J(r_{h})=2M\phi_{0}/\lambda$.

The form (\ref{fe3}) of the gravitational field equations
appears more suitable for making contact with fluid dynamics.
For this reason in the following we will consider the equations for
the gravitational dynamics in the form given by Eqs.\ (\ref{fe3}).
With this choice  the constant $\phi_{0}$  becomes irrelevant.
Therefore   we will set
$\phi_{0}=1$, and correspondingly $F_{0}=1/\la$ in Eq.\ (\ref{mass}).

The gravitational metric (\ref{bh}) can be related to the acoustic metric
associated with the  steady, cylindrically symmetric   flow of a
 barotropic, inviscid and locally irrotational 3D fluid \cite{Cadoni:2004my},
\beq\lb{acoustic}
ds^{2}= \frac {\rho_{0}}{c}\left[ -\left( c^{2} -v_{0}^{2}\right) dt^{2}
-2 v_{0} dxdt + dx^{2}\right ],
\feq
where $x$ is the coordinate along the flux tube, $\bar\rho_0(x)$ is the density
of the fluid, $v_0(x)$ its velocity and $c(x)=\sqrt{dP/d\bar\rho}$ is
the velocity of sound ($P$ is the pressure of the fluid). Owing to the
cylindrical symmetry of the flux tube the fluid parameters 
$\bar\rho,v_{0},c$,  the section of the flux tube and the potential
for the external forces acting on the fluid are independent of the
transverse coordinates $y,z$ of the flux tube.  
For
notational convenience, in (\ref{acoustic}) we use the
 dimensionless variable $\rho_0=\bar\rho_0/\la^4$.
The mapping between the gravitational (\ref{bh}) and acoustic
(\ref{acoustic}) metric  is  realized by the coordinate transformation
\bea\lb{pg}\nonumber
r&=&\int \rho_{0}dx,\quad\quad
\tau= t+ \int dx \frac{v_{0}}{ c^{2}-v_{0}^{2}},\\
J &-&\frac{2M}{\la} = \frac{\rho_{0}}{c}\left( c^{2}
-v_{0}^{2}\right).
\eea
Considering a flux tube of section  $\bar A(x)$, the dynamical
equations governing the motion of the fluid are the Euler and the
continuity equations
\bea
&&\bar\rho_0v_0{dv_0\over dx}+{dP\over dx}+\bar\rho_0{d\psi\over
dx}=0,\nonumber\\
&&{d{\cal F}
\over dx}\equiv{d\over dx}(\bar\rho_0v_0\bar A)=0,
\eea
where $\cal F$ is the flux of fluid mass and
$\psi$ is the potential for external forces acting on the fluid.

Let us now
compare the gravitational equations with the fluid-dynamical
equations obtained defining the variables
\beq\lb{nv}
X= \frac{\rho_{0}}{c}\left( c^{2}- v_{0}^{2}\right),\quad
Y= \rho_{0}c,\quad
F= \ln \left(\frac{c}{\rho_{0}}\right).
\feq

The Euler and continuity equations take, respectively, the form
\bea\lb{fd1}
&&\frac{dX}{dr}=2\frac{dY}{dr}-X\frac{dF}{dr}+
2e^{-F}\frac{d\psi}{dr},\nonumber\\
&&\frac{d}{dr}\sqrt{A^{2} Y\left( Y- X\right)}=0,
\eea
where we have introduced the
dimensionless flux tube section $A=\la^2\bar A$.

The fluid dynamic equations (\ref{fd1}) are put in the
gravitational form (\ref{fe3})  by the identification
\beq\lb{id}
X=U,\quad\quad
\sqrt{A^{2}Y(Y-X)}= \frac{M(r)}{\la},
\feq
(actually,  on the right hand side of the second equation
one could choose any function of $M(r)$),
and by introducing the constraint
\beq\lb{constraint}
2\frac{dY}{dr}-X\frac{dF}{dr}+ 2e^{-F}\frac{d\psi}{dr}=\la^{2}
V(\phi)\frac{dr}{d\phi}.
\feq
In Eq.\ (\ref{id})  the $\lambda$ factor has been introduced for
dimensional reasons.
Notice that the identification (\ref{id}) and the  constraint
(\ref{constraint}) hold both for a flux tube of constant and
non-constant section.

In terms of the solutions of the gravitational field equations
(\ref{bh}) the identifications (\ref{id}) read
\beq\lb{id1}
X=\left( J(\lambda  r)-
\frac{2M}{\lambda}\right),\quad\quad \sqrt{A^{2}Y(Y-X)}\equiv\frac{{\cal
F}}{\lambda^{2}}=
\frac{M}{\la},
\feq
where now $M$ is $M(r)$ evaluated on-shell.
The identification given by the second equation in (\ref{id1}) is very
natural and has a simple physical meaning. It identifies the conserved
quantity  of fluid dynamics, the flux of matter fluid
${\cal F}$, with the conserved quantity of the gravitational dynamics,
the black hole mass $M$.

\subsection{ Gravitational  dynamics  with pointlike sources and fluid
dynamics}

Let us  now generalize the previous results to the case $m\neq0$.
Also in presence of a pointlike source the equations (\ref{fem})
can be rewritten in terms of $M(r)$, in a form similar to (\ref{fe3}),
as

\beq\lb{fem2}
\frac{dU}{dr}\frac{d\phi}{dr}=\lambda^{2}V,\quad\quad
{dM\over dr}=\frac {m }{\lambda}W{d\phi\over dr}\,\de(r).
\feq

Introducing the variables (\ref{nv}) we find that also in this
case the identification (\ref{id}), allows us to find the acoustic
analog of the gravitational system (\ref{fem2}). The Euler
equation in (\ref{fd1}) and the constraint (\ref{constraint}) are
unchanged by the coupling to matter, and are still equivalent to
the first equation in (\ref{fem2}). The second equation in
(\ref{fem2}) implies a modification of the continuity equation in
(\ref{fd1}), which acquires a term proportional to the parameter
$m$, 
\beq \lb{fem3}
{d\over dr}(A\rho_0v)=\frac{m}{\lambda^{2}}
W\frac{d\phi}{dr}\de(r). \eeq 
This can be recognized as the
continuity equation in the presence of a flux source of strength
\beq\lb{fem4}
\Phi=\frac{m}{\lambda^{2}}\left[W(\phi)\frac{d\phi}{dr}\right]_{r=0}\
\de(r). \eeq The acoustic analog of the gravitational singularity
 is therefore simply a source
term for the flux of the fluid. Owing to the cylindrical symmetry
of our flux tube the source is planar. Using the first equation in
(\ref{pg}) one can trade in Eqs. (\ref{fem3}),(\ref{fem4})   the radial
coordinate $r$ for the the flux tube coordinate $x$. In terms of
the coordinate $x$ the source  is located at the position $x=x_0$
corresponding to $r=0$ and it extends  along the perpendicular
$y,z$ directions to cover the whole section of the flux tube.

In the gravitational description the position of the source is
associated with a singularity of the dynamics (curvature
singularity). This is not the case with the acoustic description
where  the delta function singularity at $r=0$ appears as a pure
source term for the fluid flux. The Euler equation and the
constraint (\ref{constraint}) are unaffected by the presence of
the source. We see here an important difference between the
gravitational and the  acoustic case. In general relativity the
presence of a pointlike source implies a (curvature) singularity
for the gravitational field. This is not the case for the fluid
where the presence of the source is completely disentangled from
the appearance of a singularity  in the dynamics.

The solutions of the gravitational field  equations (\ref{fem}) have
been derived in Ref.\ \cite{Cadoni:2005fw}. The delta function
singularity at $r=0$ is generated by taking the
solution as function of $|r|$,
\beq\lb{gs}
U=\frac{J(\phi)}{\sigma^{2}}-\gamma,\quad \phi=\sigma \la |r|+\beta,
\feq
where as usual $J=\int d\phi V$ and $\sigma,\beta, \gamma$ are
integration constants satisfying
\beq\lb{ic}
J(\beta)-\sigma^{2}\gamma= - \frac{m\sigma}{2\la} W(\beta),\quad
V(\beta)=-\frac{m\sigma}{\la} \frac {dW}{d\phi}(\beta).
\feq
A third equation constraining the values of the integration constants
has to be added to ensure the equality between the mass $M$ of the
solution and the mass $m$ of the source \cite{Cadoni:2005fw}):
\beq\lb{ic1}
\gamma= \frac{2m}{\la\sigma^{2}}.
\feq
Notice that, as expected,    solution (\ref{gs})   differs from the
vacuum solution (\ref{bh}) only at $r=0$. For $r>0$ the two solutions
are related by a translation of the coordinate $r$ and a scale
transformation of $r$ and $\tau$.
One can now easily work out the coordinate transformations relating
the  gravitational black hole  (\ref{gs}) to the acoustic black hole
(\ref{acoustic}). They are given by Eq.\ (\ref{pg}) with the third
equation replaced by
\beq\lb{ct}
\sigma^{-2}J-\gamma=\frac{\rho_{0}}{c}\left( c^{2}
-v_{0}^{2}\right).
\feq

\section {Black hole thermodynamics }
The correspondence between  gravitational and fluid dynamics
allows us to straightforwardly define
the thermodynamical parameters, $T_{a}, M_{a}, S_{a}$
associated with the  acoustic black hole, which satisfy the first
principle $dM_{a}=T_{a}dS_{a}$ \cite{Cadoni:2004my}.
The relevant formulas expressing the thermodynamical parameters in
terms  of the variables $X,Y,F$ of Eq.\ (\ref{nv}) have been already
given in Ref.\ \cite{Cadoni:2004my}. However, the acoustic black hole
thermodynamics developed in that paper is not completely
satisfactory. The thermodynamical parameters associated with the
acoustic black hole  are a simple ``translation'' of those
associated with the gravitational black hole. A physical
interpretation of them, in particular of the mass $M_{a}=M$,
in terms of fluid parameters is difficult.

The results derived in  the previous sections of this paper, in
particular Eq.\ (\ref{id1}), allow us to
give a more transparent ``fluid-dynamical'' interpretation of the
thermodynamical parameters associated with the acoustic black hole.
From Eq.\ (\ref{id1})  it follows that the black
hole mass $M$ is the  flux of matter fluid
measured  in units of $\lambda$,
 $M_{a}={\cal F}/\lambda$. Because $\cal{F}$ (and $M$) are constant
 of motion we can evaluate them on the horizon $r=r_{h}$. We have,
 \beq\lb{flux}
 M_{a}=M(r_{h})= \la A(r_{h})Y(r_{h)}.
 \feq
The entropy of the acoustic black hole is  \cite{Cadoni:2004my},
$S_{a}= 2\pi \la r_{h}$. For a flux tube of constant section the
entropy is proportional to the total mass of the fluid
${\cal{M}}(r_{h})$ inside the horizon.
In fact from Eq.\ (\ref{pg}) it follows
\beq
S_a=2\pi\la\int_{r=0}^{r_{h}}\rho_{0}dx=
\frac{2\pi}{\la A}{\cal{M}}(r_{h}).
\eeq

The thermodynamical parameters satisfy the first principle. Making
use of Eq.\ (\ref{flux}) we therefore  have for the black hole
temperature, \beq\lb{temp} T_{a}= \frac{dM_{a}}{dS_{a}}=
\frac{1}{2\pi} \frac{d}{dr_{h}}
\left[A(r_{h})Y(r_{h})\right]=\frac{1}{2\pi\la^{2}}
\frac{d{\cal{F}}(r_{h})}{dr_{h}}. \feq The temperature of the
acoustic black hole measures  the rate of change of the flux of
fluid  mass when the position of the horizon is changed. For a
flux tube of constant section, for which $r_{h}$ is proportional
to the mass of the fluid inside the black hole, the temperature
measures the rate of change of the flux  when the mass of the
fluid inside the hole is changed.

\section{Solutions of the constrained fluid dynamics}
In the previous sections we have shown that the gravitational dynamics
governing the spherically symmetric solution  of the action (\ref{ea})
has an acoustic analogue given by a fluid dynamics constrained by  Eq.
(\ref{constraint}).  If the  geometry of the flux tube and the
external forces are fixed, this constraint determines the equation of
state of the fluid. The physically relevant situation, which we
consider here, is the opposite.
One works with a fluid with a given equation of state and Eq.
(\ref{constraint}) becomes a constraint on  the geometry of the flux
tube and/or on the
external forces acting on the fluid.

In Ref.\ \cite{Cadoni:2004my}  the constraint
(\ref{constraint})  has been  solved for a perfect fluid.
In the following we will solve  the constrained fluid dynamics for a fluid with a
generic power-law equation of state
\beq\lb{es}
P=\la^{4}\frac{a^2}{n} \rho_{0}^{n}.
\feq
where $a$  and $n\neq 0$ are  real dimensionless constants.
This equation of state describes almost all physically  interesting
fluids: perfect fluid ($n=1$), Bose-Einstein condensate ($n=2$),
Chaplygin gas ($n=-1$), fluid mechanics in $d$ spatial dimensions
invariant under the nonrelativistic
conformal group $SO(1,2)$ ($n=1+2/d$) \cite{Jackiw:2004nm}.

In terms of the variables (\ref{nv}) the equation of state reads
\beq\lb{es1}
Y^{3-n}e^{(n+1)F}=a^{4}.
\feq
The particular cases $n=1$ (perfect fluid), $n=-1$ (Chaplygin gas)
$n=3$  ( conformal invariant fluid mechanics in $d=1$ spatial
dimensions), corresponding
respectively to  $Ye^{F}=const$, $Y=const$, $F=const$ will be
considered separately.

Using Eq.\ (\ref{es1}) in (the inverse of)  Eqs.\ (\ref{nv}) we find
the fluid velocity, the fluid density and the speed of sound
\beq\lb{es2}
v_{0}=a^{\frac{2}{n+1}} Y^{\frac{n-3}{2(n+1)}}\sqrt{Y-X},
\feq

\beq\lb{es2a}
 \rho_{0}=a^{-\frac{2}{n+1}} Y^{\frac{2}{n+1}},\quad
c=a^{\frac{2}{n+1}} Y^{\frac{n-1}{n+1}}.
\feq
We will discuss separately the two cases of a flux tube
of constant and non-constant section.
\subsection{Flux tube with constant section}

In this case $A= const$ and  we can use the continuity equation to solve for
$X=X(Y)$:
\beq\lb{f1}
X=Y-\frac{\alpha^{2}}{Y},
\feq
where $\alpha= M/(A\la)$.
Using Eq.\ (\ref{f1}) into Eq.\ (\ref{es2}) we can express also $v_{0}$ as a
function of $Y$:
\beq\lb{f3}
v_{0}= \alpha\left({a\over Y}\right)^{\frac{2}{n+1}}
\feq
The constraint (\ref{constraint}) can be written as
\beq\lb{con1}
\frac{dX}{dY}-2 +X\frac{dF}{dY}-2 e^{F}\frac{d\psi}{dY}=0,
\feq
which using  Eq.\ (\ref{es1}) can be easily solved for $\psi(Y)$,
\beq\lb{f4}
\psi= -a^{\frac{4}{n+1}}\left(\frac{\alpha^{2}}{2}
Y^{-\frac{4}{n+1}}+\frac{1}{n-1}
Y^{2\frac{n-1}{n+1}}\right).
\feq
The  acoustic horizon forms at $Y=\alpha$, corresponding to $X=0$.
The  subsonic region ($X>0$), describing the external region of
the gravitational black hole, is given by $Y>\alpha$.  The source,
corresponding to the  black hole singularity,  is located at
$Y=Y_{s}$ with  $ 0<Y_{s}<\alpha$. The fluid parameters
$v_{0},\rho_{0}, c$ and the external potential $\psi$ remain
finite both on  the horizon and on the location of the source. In
the  case under consideration  the constraint necessary to have a
correspondence between gravitational and fluid dynamics takes the
form of a  constraint on the form of the external forces acting on
the fluid, given by Eq.\ (\ref{f4}). Moreover, this constraint
implies a ``null force condition'' on the acoustic horizon. In
fact  the external potential $\psi$ given by Eq.\ (\ref{f4})
becomes  extremal for $Y=\alpha$ (a local maximum for $n>-1$  and
a local minimum for for $n<-1$).

Once a specific action for the four-dimensional  gravity model
(\ref{ea}), and  hence a specific $V(\phi)$ in the 2D model (\ref{dg}), is
given,
one can use  Eq.\ (\ref{f1}) to find $Y(r)$. Use of  Eqs.\ (\ref{es2a}),
(\ref{f3}) and (\ref{f4}) allows one  to get the fluid parameters as a function
of $r$.
\subsection{Flux tube with non-constant section}

In this case $A$ is not constant and
we can use the continuity equation to
express it as function of $X$ and $Y$,
\beq\lb{se}
A^{2}= \frac{\alpha^{2}}{Y(Y-X)}.
\feq
The presence of
external forces is now an unnecessary complication, and therefore we set them
to zero. With $\psi=const$, and using Eq.\ (\ref{es1}) the constraint
(\ref{constraint}) becomes
\beq\lb{con3}
2\frac{dY}{dX}-\frac{n-3}{n+1}\frac{X}{Y}\frac{dY}{dX} =1,
\feq
which is readily solved to give
\beq\lb{f5}
X= \frac{n+1}{n-1}\left(Y-\omega^{2}\, Y^{-\frac{n-3}{n+1}}\right),
\feq
where $\omega$ is an integration constant.
Inserting the previous equation into Eqs.\ (\ref{es2}), (\ref{es2a})
and (\ref{se}), we get the solution of the constrained fluid dynamics.
The acoustic horizon forms at $Y=Y_{h}= \omega^{(n+1)(n-1)}$ and
the source is located at $Y=Y_{s}$, with  $0<Y_{s}<Y_{h}$.
The supersonic region, corresponding to the black hole
interior, is given by  $Y_{s}<Y<Y_{h}$. Conversely, the subsonic region (black hole
exterior) is given by $Y>Y_{h}$.
The fluid parameters
$v_{0},\rho_{0}$ and the speed of sound $c$  remain  finite
both at   the horizon and at the location of the source.

We can also show that for $n>-1$ the acoustic horizon always forms
at a minimum of the section $A(x)$, i.e.\ the flux tube must have
the form of a so-called  Laval nozzle. In fact it follows from
Eq.\ (\ref{se}) and  Eq.\ (\ref{con3}) that
\beq\lb{f4a}
\frac{dA}{dX}=\frac {4 X }{Y-X}\frac{1}{2(n+1)Y-(n-3)X}. \feq
Taking into account that by definition $Y>0$ and that  Eq.\
(\ref{se}) requires $Y >X$, one finds that for $n>-1$,  $dA/dX=0$
at the horizon ($X=0$), whereas it is positive (negative) for
$X>0$ ($X<0$). This is a highly  non-trivial result. The
constraint, which is necessary to have a correspondence between
gravitational and fluid dynamics becomes  a geometrical constraint
on the form of the flux tube. This  geometrical constraint  forces
the fluid to develop an acoustic horizon.

\subsection{Particular cases: $n=1, -1,3$}
The case $n=1$ corresponding to a perfect fluid has been already discussed
in Ref.\ \cite{Cadoni:2004my}.
For $n=-1$ we have the Chaplygin gas. The pressure is negative and the
equation of state becomes  $P= -\la^{4}a^2/\rho_{0}$, which in terms of
the new variables (\ref{nv}) reads $Y=a$. In the case of a flux tube of
constant section, constant $Y$ implies, owing to the continuity
equation, also constant $X$. This forbids the realization of an
acoustic Chaplygin  analogue of  a gravitational black hole,
independently of the form of the potential $V$.
This difficulty can be circumvented by considering  a flux tube of
non-constant section. In this case, considering for simplicity $\psi=0$,
the constraint (\ref{constraint}) can be solved to give  $\exp(-F)=
bX$, with  integration constant $b>0$.  The fluid parameters  take
therefore the form
\beq\lb{f26}
\rho_{0}=  \sqrt{abX},\quad v_{0}=
\sqrt{\frac{1}{b}\left(\frac{a}{X} -1\right)},\quad
c=\sqrt{\frac{a}{bX}}, \quad A^{2}= \frac{\alpha^{2}}{ a^2-aX}.
\eeq
This solution is defined only for  $X\le a$ and  becomes   singular
on the black hole horizon $X=0$.

For $n=3$  the equation of state implies $F=const$ and  the
solutions take a very simple form. In the case of a flux tube of
constant section Eqs. (\ref{es2a}),(\ref{f1}), (\ref{f3}),
(\ref{f4})  give
\bea\lb{f7} v_{0}&=& \alpha
\left(\frac{a}{Y}\right)^{1/2},\quad \rho_{0}=\left(\frac{a}{Y}
\right)^{-1/2},\quad
c=(aY)^{1/2},\\
Y&=& \frac{1}{2}\left( X-\sqrt{X^{2}+4\alpha^{2}}\right),\quad  \psi=
-\frac{a}{2} \left(\frac{\alpha ^{2}}{Y}+Y \right).
\eea

In the case of a non-constant flux tube section we get instead
\beq\lb{f8} v_{0}=  \sqrt{a\left(d -\frac{X}{2}\right)},\quad
\rho_{0}= \sqrt{\frac{1}{a}\left(d +\frac{X}{2}\right)}\quad
c=a\rho_{0},\quad  A= \frac{\alpha}{v_{0}c}. \feq where $d>0$ is
an integration constant related to the integration constant
$\omega$ appearing in Eq.\ (\ref{f5}). It is evident from the
previous formulae that $-2d\le X\le 2d$. The horizon is located at
$X=0$, which is also a minimum of  $A$. The section of the flux
tube diverges at $X= \pm 2 d$. Therefore the acoustic black hole
cannot   describe the entire black hole spacetime.

\section{conclusions}
The resolution of the singularities of general relativity relies
heavily on
the understanding of the short distance, quantum  behavior of gravity.
Nevertheless, investigations of  singularities  at the classical
level may also  help  to improve our understanding of the subject.
In this paper we have presented acoustic analogues of the
black hole singularities generated by a pointlike source.
The acoustic analogues of the singularities are constructed finding a
fluid whose dynamics is completely equivalent to that governing the
spherically symmetric solutions of (gauge fixed)
general relativity with pointlike sources.

This result enhances the analogy between gravity and
condensed matter systems. This analogy is not restricted to
kinematical features of gravity such as acceleration horizons.
We can use a fluid to mimic the full black
hole dynamics, spacetime singularity included.
Analogue models of gravity
have also the nice feature of being, at least in principle,
experimentally testable in laboratory. In the
near future this could open the way to  experimental tests of the full
classical black hole  structure realized using condensed
matter systems.

Another important result of our investigation concerns the notion of
singularity and source in classical field theory and, in particular,
in general relativity.
In a classical field theory, such as Maxwell electromagnetism, the
(delta function) singularity associated  with
a pointlike  source is  conceptually independent from the  field
singularity, the point where the fields diverges.
It is the classical dynamics (e.g.\ the Laplace equation)
that forces the identification of the two types of singularities.
The delta function singularity becomes  the source of the Green
function for the field. In general relativity this relationship
between pointlike sources and field singularities becomes even deeper.
The equivalence principle treats on equal footing sources and
test particles, and the field singularities become geometrical
(curvature singularities).

In this paper we have shown that for a classical field theory
describing the motion of a fluid we may have a slightly different
situation.  The presence of a  source for the mass of the
fluid is not necessarily related to a divergence of the fields
describing the fluid. It is related with a much milder form of singularity,
a cusp point in the field configuration.
This result is consequence of two  features
of the the fluid dynamics  we have considered. First the effective theory
which describes the fluid  is two-dimensional. In two dimensions the delta-like
singularity of the source can be generated  by
a second derivative of $|r|$ in the Green equation.
Second, differently  from other field
theories, in fluid dynamics   the source
couples, through the continuity equation,   to the
flux of matter fluid only.

By mapping gravitational and fluid dynamics,  we have also shown
that the singularity of spherically
symmetric black hole spacetime generated by a pointlike source can be
transformed in the  milder singularity of fluid dynamics   described above.
The responsible for this miracle is the 2D dilaton-dependent Weyl
rescaling of the
metric (\ref{wr}). The price that we have to pay for this  is
that the Weyl transformation itself is singular at the position of
the source  (see eq.\ (\ref{gs})). The Weyl transformation changes
the  the 4D black hole singularity into a 2D cusp singularity.

In this paper we have considered only spherically symmetric solutions
of general relativity. It is an interesting open question if our
results hold also in more general situations. It seems very difficult
to mimic a  gravitational dynamics with  propagating degrees of
freedom or a system with two or more interacting masses,  using a fluid.
Nevertheless a description similar to that
presented in this paper may still hold for classes of solutions with
high degree of symmetry, such as for instance cosmological solutions.


\end{document}